\documentclass[conference]{IEEEtran}
\IEEEoverridecommandlockouts
\usepackage{balance}
\usepackage{epsfig,endnotes}
\usepackage{import}
\usepackage{url}            
\usepackage{commath}
\usepackage{booktabs}       
\usepackage{amsfonts}       
\usepackage{nicefrac}       
\usepackage{microtype}      
\usepackage{multirow}
\usepackage{graphicx}
\usepackage{xcolor}
\usepackage{subcaption}
\usepackage{array}
\usepackage[hidelinks]{hyperref}
\usepackage{cite}
\usepackage{amsmath,amssymb,amsfonts}
\usepackage{algorithmic}
\usepackage{textcomp}
\def\BibTeX{{\rm B\kern-.05em{\sc i\kern-.025em b}\kern-.08em
    T\kern-.1667em\lower.7ex\hbox{E}\kern-.125emX}}

\usepackage[linesnumbered,ruled,vlined]{algorithm2e} 

\SetKwInput{KwInput}{Input}                
\SetKwInput{KwOutput}{Output}              

 \author{\IEEEauthorblockN{Giulio Pagnotta}
 \IEEEauthorblockA{Dipartimento di Informatica\\
 Sapienza University of Rome\\
 pagnotta@di.uniroma1.it}
 \and
 \IEEEauthorblockN{Dorjan Hitaj}
 \IEEEauthorblockA{Dipartimento di Informatica\\
 Sapienza University of Rome\\
 hitaj.d@di.uniroma1.it}
 \and
  \IEEEauthorblockN{Fabio De Gaspari}
 \IEEEauthorblockA{Dipartimento di Informatica\\
 Sapienza University of Rome\\
 degaspari@di.uniroma1.it}
 \and
 \IEEEauthorblockN{Luigi V. Mancini}
 \IEEEauthorblockA{Dipartimento di Informatica\\
 Sapienza University of Rome\\
 mancini@di.uniroma1.it}}

\begin{document}

\title{PassFlow: Guessing Passwords with Generative Flows}

\maketitle
\begin{abstract}
Recent advances in generative machine learning models rekindled research interest in the area of password guessing. Data-driven password guessing approaches based on GANs, language models and deep latent variable models have shown impressive generalization performance and offer compelling properties for the task of password guessing.

In this paper, we propose PassFlow, a flow-based generative model approach to password guessing. Flow-based models allow for precise log-likelihood computation and optimization, which enables exact latent variable inference. Additionally, flow-based models provide meaningful latent space representation, which enables operations such as exploration of specific subspaces of the latent space and interpolation. We demonstrate the applicability of generative flows to the context of password guessing, departing from previous applications of flow-networks which are mainly limited to the continuous space of image generation. 
We show that PassFlow is able to outperform prior state-of-the-art GAN-based approaches in the password guessing task while using a training set that is \emph{orders of magnitudes smaller} than that of previous art.
Furthermore, a qualitative analysis of the generated samples shows that PassFlow can accurately model the distribution of the original passwords, with even non-matched samples closely resembling human-like passwords.

\end{abstract}


\thispagestyle{empty}
\section{Introduction}
\label{sec:introduction}

Several decades after their introduction in the field of computer science, text-based passwords continue to be the most widely used authentication mechanism. Password-based systems are simple to implement, efficient and familiar to users. However, passwords suffer from well-known drawbacks and vulnerabilities, mainly due to the limited complexity and inherent structure present in human-generated passwords, which heavily restrict the regions of space in which such passwords reside. Traditional password guessing tools, such as John The Ripper(JTR)~\cite{jtr} and HashCat~\cite{hashcat}, exploit this markedly uneven distribution in the password space to generate high-probability password guesses that fall in the dense areas of the space where human-like passwords reside. These tools are able to approximate the distribution of human-like passwords primarily based on carefully generated rules handcrafted by human experts, which is a laborious task that requires a high level of domain-specific expertise.

To address the limitations of traditional tools, in recent years different unsupervised learning-based approaches to password guessing based on generative models have been proposed. These generative models are carefully designed to autonomously learn structure and patterns that are characteristic of human-generated passwords, with the goal of improving password guessing performance and removing the need for domain-specific expertise. Generative Adversarial Network (GAN) based approaches, such as PassGAN~\cite{passGAN} and Pasquini \textit{et al.}~\cite{pasquini_representation_learning}, are able to learn an implicit representation of the password space based on an approximation of the probability density function of the generation process. GAN-based approaches are therefore unable to directly provide estimates for the probability density of a sample, but are still able to generate high probability passwords that are very close to human generated passwords. Moreover, GANs cannot directly represent samples in a latent space unless an encoder is separately trained, and might not have full support over the data distribution~\cite{glowINN}.
Password guessing approaches based on deep latent variable models, such as Variational Autoencoders (VA) and Wasserstein Autoencoders~\cite{pasquini_representation_learning} (WAE), explicitly learn a mapping between the password distribution and a prior latent space distribution based on the Evidence Lower Bound (ELBO) loss function. Differently from GANs, these models provide a direct way to evaluate the probability density of a sample and can also generate high probability passwords that follow the distribution of human generated passwords. However, due to the reliance on the ELBO function for optimization, these models only learn an approximation of the posterior probability $p(z|x)$ and, therefore, an approximation of the latent variable $z$ corresponding to a datapoint $x$.

This paper introduces \textit{PassFlow}, a novel password guessing approach based on generative flows~\cite{dinh2014nice}. Generative flows provide several advantages over GANs and deep latent variable models.
First, differently from GANs, flow networks provide an explicit latent space by learning an invertible mapping between a datapoint $x$ and its latent representation $z$. The availability of an explicit latent space enables complex operations on latent points $z$ such as interpolation between different points and meaningful alterations of existing points.
Second, differently from VA and WAE, flow networks are optimized using exact log likelihood computation, rather than with a function providing a lower bound for the expectation of a sample. This means that flow models are able to provide exact latent variable inference, compared to an approximate inference for VA and WAE. 
We show that these properties allow flow-based generative models to outperform deep latent variable models in the password guessing task, while remaining competitive with state-of-the-art GANs. However, while GAN architectures have seen extensive research and fine-tuning, flow-based models have just began to attract significant attention from the research community~\cite{realNVP,glowINN,MaCow,NEURIPS2019_5d0d5594}. It seems reasonable that, in the near future, flow-based architectures will also see extensive improvements and fine-tuning. Moreover, flow networks enable latent space operations such as interpolation which are not directly possible with GAN architectures, unless separate models are trained for encoding.

To summarize, our contributions are as follows:
\begin{itemize}
       \item We introduce PassFlow, a flow-based generative architecture for password guessing, proving its applicability in this domain. To the best of our knowledge, this is the first work that shows the applicability of flow networks to the password guessing problem, as well as one of the first works exploring their use in domains other than image generation.
       
      \item We extensively evaluate PassFlow and show that it outperforms both state-of-the-art GANs and deep latent variable models in the password guessing task.
       
       \item We show PassFlow can generalize extremely well with a dataset that is \emph{two orders of magnitude smaller} compared to state-of-the-art GANs and deep latent variable models. 
     
       This means that, contrary to prior art architectures, PassFlow can be effectively used even when only a small subset of the target password set is known.

       \item We thoroughly analyze the structure of the latent space learned by PassFlow and show that it is able to capture a meaningful representation of the data space, allowing operations such as interpolation between different password samples and exploration of dense areas of the space.

\end{itemize}

This paper is organized as follows: in Section~\ref{sec:background} we present relevant background knowledge; in Section~\ref{sec:passflow} we introduce PassFlow, our flow-based generative approach for password guessing; in Section~\ref{sec:evaluation} we present our experimental results. Section~\ref{sec:related_work} discusses the related work in the password guessing domain and finally Section~\ref{sec:conclusions} discusses future work and concludes the paper.
 
\section{Background}\label{sec:background}

Flow-based generative models are a category of deep latent variable models based on the concept of flows~\cite{dinh2014nice,realNVP}. A flow is a bijective function $f_\theta$ from a domain $X \subseteq \mathbb{R}^D$ to a codomain $Z \subseteq \mathbb{R}^D$, where $\theta$ represent the parameters learned by the model. Typically, in flow networks $f_\theta$ is a composition of a sequence of k bijective functions $f_i$, such that the relationship between a random variable $x \in X$ and its latent representation $z \in Z$ is as follows:

\begin{align}
\label{eq:z_flow}
z = f_\theta(x) = (f_k \circ f_{k-1} \circ ... \circ f_1) (x)
\end{align}

Given that each $f_i$ is bijective, $f_\theta$ is also bijective and x can be obtained given a latent representation z simply by:

\begin{align}
\label{eq:inv_f}
x = f_\theta^{-1}(z) = (f_1^{-1} \circ f_{2}^{-1} \circ ... \circ f_k^{-1}) (z)
\end{align}

Under the change of variable formula, using Equation~\ref{eq:inv_f}, we can define the probability density function $p_\theta(x)$ as follows:

\begin{align}
\label{eq:pdfx}
p_\theta(x) &= p_z(z) \abs{det \left( \frac{\partial z}{\partial x} \right)} \\
\label{eq:pdfx_full}
&= p_z(f_\theta(x)) \abs{det \left( \frac{\partial f_\theta(x)}{\partial x} \right)} \\
\label{eq:log_pdfx}
log(p_\theta(x)) &= log(p_z(f_\theta(x))) + log \left( \abs{det \left( \frac{\partial f_\theta(x)}{\partial x} \right)} \right) \\
&= log(p_z(f_\theta(x))) + \sum_{i=1}^{k} log \left( \abs{det \left( \frac{\partial f_i(x)}{\partial x} \right)} \right)
\end{align}
 
\noindent
where $\frac{\partial f_\theta(x)}{\partial x}$ is the Jacobian matrix of the function $f_\theta$ at point x and $p_z$ is an easy-to-sample, factorized prior distribution for z (e.g., multivariate gaussian). In this context, the Jacobian matrix can be thought of as a scale parameter that defines the change in (log-)density when going from x to $f_\theta(x)$.

Flow models are typically trained by minimizing the negative log-likelyhood function. Given a dataset of samples $I = \{x_i | 0 \leq i < N\}$, the loss function is defined as:

\begin{align}
\label{eq:loss}
\mathcal{L}(I) &= \frac{1}{N} \sum_{i=1}^{N} - log(p_\theta(x_i))    \\
\label{eq:full_loss}
&= \frac{1}{N} \sum_{i=1}^{N} - log(p_z(f_\theta(x_i))) - log \left( \abs{det \left( \frac{\partial f_\theta(x_i)}{\partial x_i} \right)} \right)
\end{align}

\noindent
where Equation~\ref{eq:full_loss} is obtained by substituting Equation~\ref{eq:log_pdfx} in~\ref{eq:loss}. 
In order for Equation~\ref{eq:full_loss} to be tractable, the determinant of the Jacobian matrix must be easily computable and therefore $f_\theta$ cannot be an arbitrary (invertible) function. The specific function adopted in this work and its Jacobian determinant are discussed in Section~\ref{sec:coupling_layer}.

Exact samples from the distribution $X$ can be easily calculated by sampling from the latent prior distribution $p_z$.
A sample $z \sim p_z$ is drawn from the latent distribution, and its preimage is computed with the inverse flow $x = f_\theta^{-1}(z)$. Similarly, the probability density for a datapoint x can be computed by calculating the probability of its image  $p_z(z) = p_z(f_\theta(x))$ and multiplying it by the determinant of the Jacobian of $f_\theta$ calculated at point x (Equation~\ref{eq:pdfx}).
\section{PassFlow}\label{sec:passflow}

This section introduces \textit{PassFlow}, a novel application of invertible flows in the domain of password guessing. Invertible flows have shown promising results in the image domain, where they have been successfully used to model the continuous distribution of natural images, generating realistic samples and enabling complex latent space manipulations~\cite{dinh2014nice,realNVP,pasquini_reducing_bias}. Inspired by these recent advances, we modify and adapt these architectures to work in the 1-dimensional domain of passwords.  To the best of our knowledge, we are the first to adopt and prove the effectiveness of flow architectures in this domain, and one of the first to explore the applicability of flow networks outside of the continuous domain of image generation~\cite{pmlr-v97-ziegler19a}. 

We base our architecture on the work of Dinh \textit{et al.}~\cite{realNVP} to learn a model that is bijective and has tractable Jacobian determinant, allowing exact negative log-likelihood optimization.

\subsection{Architecture of the Coupling Layers}
\label{sec:coupling_layer}
As discussed in Section~\ref{sec:background}, a critical requirement for the architecture of the network is that the Jacobian determinant is tractable and that the inverse function $f_\theta^{-1}$ is easily computable. The coupling layer is the component of the invertible neural network architecture that ensures these properties. Each coupling layer is roughly equivalent to a single step $f_i$ of the function composition $f_\theta$ (see Equation~\ref{eq:z_flow}). Our work bases the design of the coupling layers on the work of Dinh et al.~\cite{realNVP}, where the authors propose the use of a coupling layer that has the following structure:

\begin{align}
\label{eq:z1}
z_{1:d} &= x_{1:d}  \\
\label{eq:z2}
z_{d+1:D} &= x_{d+1:D} \odot exp(s(x_{1:d})) + t(x_{1:d})
\end{align}

\noindent
where : is the slicing operator, D is the dimensionality of $X$ (and, therefore, $Z$), $\odot$ is the Hadamard product and $s$ and $t$ are scale and translation functions respectively. In practice, we implement $s$ and $t$ as two residual block-based neural networks due to the impressive generalization performance of these architectures~\cite{He_2016_CVPR}. The Jacobian of the above equation can be calculated as:

\begin{align}
    \mathbb{J} = \frac{\partial z}{\partial x} = 
    \begin{bmatrix}
    \mathbb{I}_d & \quad  &0\\
    \frac{\partial z_{d+1:D}}{x_{1:d}} & \quad & diag(exp \left[ s(x_{1:d}) \right])
    \end{bmatrix}
\end{align}

 Since $\mathbb{J}$ is a triangular matrix, its determinant can be efficiently computed as the product of the diagonal elements of the matrix:

\begin{align}
    det(\mathbb{J}) =  \prod_j exp \left[ s(x_{1:d})_j \right] = exp\left[ \sum_j s(x_{1:d})_j \right]
\end{align}

The inverse of Equations~\ref{eq:z1} and~\ref{eq:z2} can also be easily computed, which means that sampling from the latent distribution and computing the preimage can be done efficiently.

\subsubsection{Masking}
\label{sec:masking}
The effectiveness of the coupling layer described above is predicated on an appropriate splitting of the input features x. Intuitively, Equation~\ref{eq:z2} is computing z by conditioning half the values of x ($ x_{d+1:D}$) on a (learned) transformation of the remaining half ($x_{1:d}$). It is therefore fundamental to chose a partitioning scheme that can effectively exploit the correlation between different elements of the input vector x.

Since passwords are the domain of our model, we use a partitioning scheme that exploits the local correlation between consecutive runs of characters. More specifically, for each layer, we define a \emph{binary mask} $b \in \{0,1\}^D$ with runs of \emph{m} consecutive zeroes and ones that is used to partition vector x:

\begin{align}
    z = b \odot x + (1-b) \odot (x \odot exp(s(b \odot x)) + t(b \odot x))
\end{align}

\begin{figure}[t]
    \centering
    \includegraphics[width=.9\columnwidth]{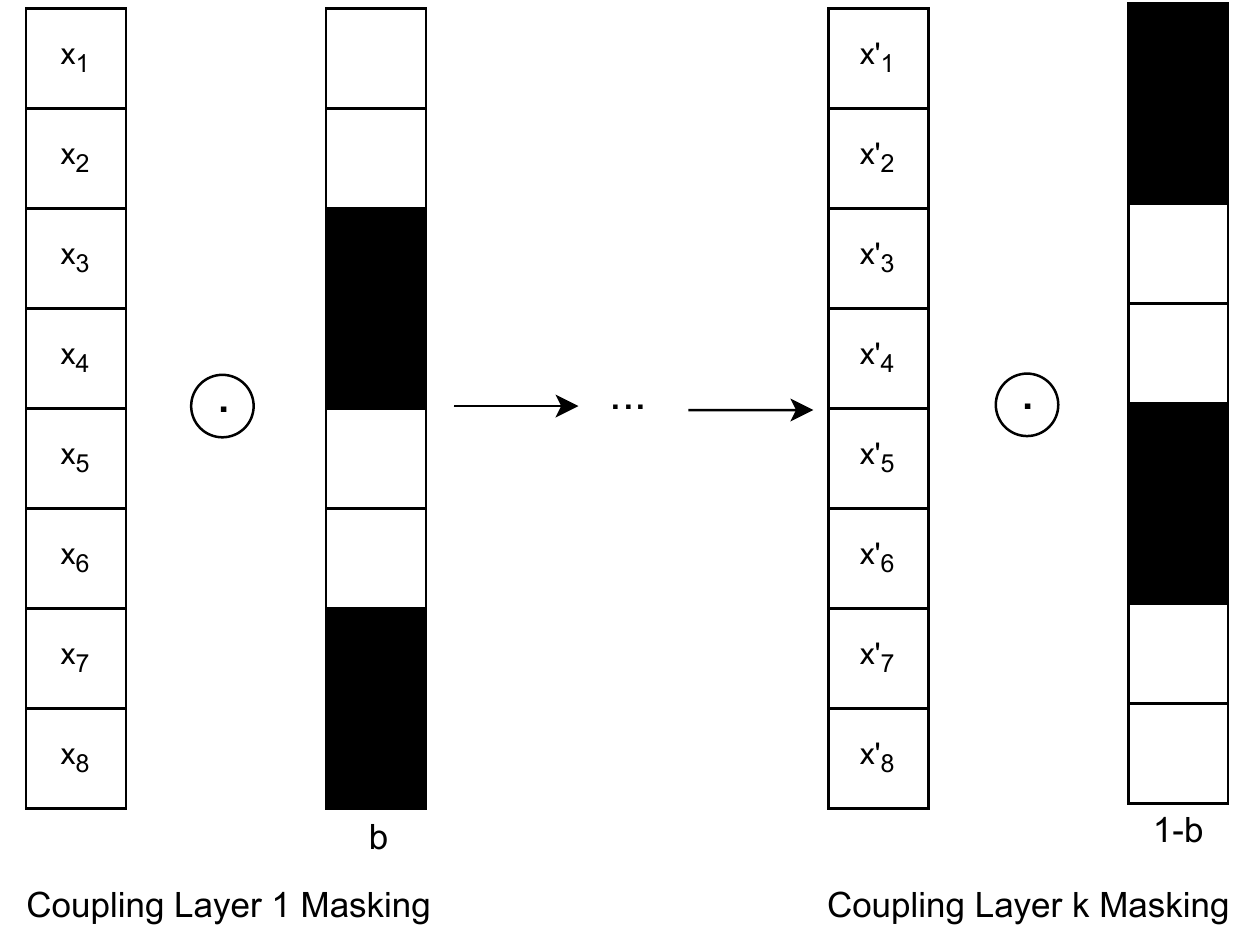}
    \caption{Example of coupling layer masking with the binary mask obtained for m=2. Binary mask $b$ and $1-b$ are alternated. The first coupling layer uses mask $b$, a later coupling layer uses the negated mask $1-b$. }
    \label{fig:masking}
\end{figure}

In order to capture complex correlations between different portions of the input x, different coupling layers of the network can use different binary masks. In particular, to avoid portions of the input being unchanged when going from $X$ to $Z$, alternating binary masks should be used in the coupling layers (see Figure~\ref{fig:masking}).

\subsection{Dynamic Sampling with Penalization}
\label{sec:ds}
Optimizing the network under the loss defined in Equation~\ref{eq:loss} forces the model to learn a non-linear transformation $f_\theta$ that maps the probability distribution $p_x$ of the domain $X$, into a (predefined) target probability distribution $p_z$ for the codomain $Z$. Given that $f_\theta$ is bijective, we can then easily generate an observation $x \in X$ by sampling $z \sim p_z$ and applying $f_\theta^{-1}$. Due to the optimization under maximum likelihood, the transformation $f_\theta$ is encouraged to map regions of $X$ with high probability density to regions with relatively high probability density in $Z$. This contraction effect is balanced by the Jacobian determinant, which penalizes contraction and favours expansions in regions of high density in the data space. 

This behaviour of $f_\theta$, coupled with the factorized structure of the prior $p_z$, has the effect that high density regions in the data space are projected to regions of the latent space with smoother density~\cite{bengio2013better}. 

The smoothness of the latent space results in a number of interesting properties. In particular, given the latent representation of data point $z_i = f_\theta(x_i)$, we find that by moving in the neighbourhood of $z_i$ we remain in a high density region of the latent space, i.e., neighbouring points of $z_i$ have a high probability of corresponding to valid samples in the data space (in our case, realistic passwords). We empirically demonstrate the smoothness of the latent space of our model in Section~\ref{sec:eval_smoothness}. We can exploit this smoothness property to greatly improve our guessing strategy. We can notice in Equation~\ref{eq:pdfx_full} that, in our generative process, the probability distribution $p(x)$ is dependent on the parameters learned by our model ($\theta$ in the equation) and on the prior distribution that we impose on our latent space ($p_z$). We can therefore directly influence the probability density that our model assigns to a point $x_i$ by altering the prior probability distribution $p_z$. This property allows us to dynamically adapt the distribution modeled by our network and condition it on the set of passwords successfully matched, by choosing a prior distribution $p_z$ that maximizes the likelihood of these matches. The resulting probability density function of the latent space is as follows:

\begin{align}
    \label{eq:ds}
    p_z(z | M) = \sum_{i=0}^{|M|} \phi(z_i) \mathcal{N}(z_i, \sigma_i)
\end{align}

\noindent
where $M$ is the set of latent points corresponding to the currently guessed passwords, $\sigma$ is the variance of the i-th Gaussian and $\phi$ is a function defining a penalization factor for the i-th matched sample in the mixture of Gaussians. The above property is also used in~\cite{pasquini_representation_learning}, however, in their work the authors use a uniform distribution to weight each sample $z_i$ in the mixture of Gaussians, while we introduce a dynamic penalization function that reduces the influence of a subset of $z_i \in M$ in the mixture over time. In particular, the coefficient $\phi(z_i)$ decreases the longer an individual matched sample $z_i$ has been in the guessed set $M$, reaching $0$ after a set number of iterations has elapsed. We further discuss the importance of the Dynamic Sampling  algorithm and the penalization factor $\phi$ in Section~\ref{sec:eval_smoothness} and ~\ref{sec:phi_train_size_performance} respectively, and their implementation in Section~\ref{sec:implementation}.

\subsection{Reducing Collisions with Data-space Gaussian Smoothing}
\label{sec:gs}
PassFlow learns a mapping for passwords between the discrete data space $X \subseteq \mathbb{N}^D$ and the continuous latent space $Z \subseteq \mathbb{R}^D$. Consequently, the model can only learn an approximation of a bijection between $X$ and $Z$, since $|\mathbb{R}| > |\mathbb{N}|$. This implies that different latent points $z_i \in Z$ are mapped to the same data point $x \in X$, which means that multiple, separate latent points when sampled can result in the same generated password (collision). In general, from our evaluation we find that the probability of a collision is higher when the sampled points are close in the latent space. This behavior introduces a tradeoff between different values of the parameter $\sigma_i$ used in dynamic sampling. 
Using large $\sigma_i$ values in Equation~\ref{eq:ds} increases the search space around a given matched sample $z_i$. This results in passwords that are potentially very different from the original matched password $x_i$ and less collisions, but lowers the chance of a successful match. On the other hand, a small $\sigma_i$ restricts the search to regions very close to $z_i$, resulting in passwords that are structurally similar to the matched password $x_i$ and increasing the probability of a successful match. However, small values of $\sigma_i$ increase the probability of collisions considerably. 

We address the problem of collisions by introducing \emph{data-space Gaussian Smoothing} (GS). After generating a data point $x_i$ by sampling and inverting a latent point $z_i$, we incrementally add small random perturbations sampled from a Gaussian distribution. By maintaining the variance of the Gaussian small, we can alter the generated data point and reduce the probability of collisions, while remaining in the neighborhood of the original point $x_i$. We show in the evaluation that data-space Gaussian Smoothing is extremely effective in preventing collisions, especially when using dynamic sampling with low $\sigma_i$ values, resulting in much higher number of matches compared to the base dynamic sampling approach.

\section{Implementation Details}
\label{sec:implementation}

This section briefly describes some details of our implementation with respect to the dynamic sampling algorithm, the penalization function $\phi$, and the interpolation algorithm used.

\subsection{Dynamic Sampling algorithm}

Algorithm~\ref{alg:ds} describes our Dynamic Sampling (DS) approach. The notation is as follows: $\Omega$ is the test set of passwords that we are trying to guess. $P$ is the set of all the passwords generated by our model that match a password in $\Omega$. The parameter $\alpha$ defines how many matches are required before dynamic sampling is activated. We use the original prior distribution $p_z$ for the latent space until a predetermined number of matches is found, after which the prior distribution is changed to that defined in Equation~\ref{eq:ds}. The rationale behind the $\alpha$ parameter is that we do not want to heavily condition the distribution probability used for sampling before an appropriate number of matches is found, as doing so could lead the generator to regions of the latent space with low probability density. $M$ is the set of latent points $z_i$ corresponding to each matched datapoint $p_i \in P$. $M_h$ is a dictionary defining, for each latent point $z_i \in M$, for how many iterations $z_i$ has been used to condition the latent space probability distribution. It is used by the function $\phi$ to calculate the penalization factor for a previously matched latent point, adding an additional dynamic factor to the definition of the latent space posterior distribution.

\begin{algorithm}[t]
\DontPrintSemicolon
  
  \KwInput{Set: $\Omega$, Int: $\alpha$, Int: $num\_guesses$}
  \KwOutput{Set: $P$}
  \KwData{Set: $P$, Set: $M$, Dict: $M_h$}

    $p_{prior}  \sim \mathcal{N}(\mu,\,\sigma)$ \;
    
   \While{$num\_guesses > 0$}
   {
        $num\_guesses \leftarrow  num\_guesses - 1$ \;
        
        $z  \leftarrow  p_{prior}$ \;
        
   		$p  \leftarrow  f_\theta^{-1}(z)$ \;
   		
   		\If{$p  \in  \Omega$} 
   		{
   		   $P  \leftarrow  P \cup  p$ \;
   		   
   		   $M  \leftarrow  M \cup  z$ \;
   		   
   		   $M_h[z] \leftarrow M_h[z] + 1$ \;
   		   
   		   \If{$|M| > \alpha$}
   		   {
   		     $p_{prior} \leftarrow \sum_{i=1}^{|M|} \phi(M_h[M_i]) \mathcal{N}(M_i,\,\sigma_i)$ \;
   	
   		   }
   		}
   }

\caption{Dynamic Sampling with Penalization}
\label{alg:ds}
\end{algorithm}

\subsection{$\phi$ function and Dynamic Sampling parameters}
\label{app:phi}

In our evaluation we implement the penalization function $\phi$ as a step function: given a threshold $\gamma$ and a matched latent point $z_j$, $\phi(z_j, \gamma)$ returns 1 if $z_j$ has been used less than $\gamma$ times in the mixture of Gaussians, and 0 otherwise.

Table~\ref{tab:parameters_ds} reports the $\alpha$, $\sigma$ and $\gamma$ values used to obtain the best performance reported in Table~\ref{table:results_comparisons_latent}. Parameter $\alpha$ represents the threshold parameter after which dynamic sampling is enabled. $\sigma$ represents the standard deviation used for each element of the mixture of Gaussians and $\gamma$ represents the threshold used in the step function.
  
\begin{table}[t]
\centering
\caption{The dynamic sampling parameters used to obtain the number of matches reported in Table~\ref{table:results_comparisons_latent}}
\label{tab:parameters_ds}
\begin{tabular}{l | c  c  c}
\toprule
{Guesses} &  $\alpha$ &  $\sigma$ & $\gamma$ \\
\midrule 
$10^4$ & 1 & 0,12 & 2  \\
$10^5$ & 1 & 0,12 & 2 \\
$10^6$ & 5 & 0,12 & 2  \\
$10^7$ & 50 & 0,12 & 10 \\
$10^8$ & 50 & 0,15 & 10 \\

\bottomrule 
\end{tabular}

\end{table}

\subsection{Interpolation algorithm}

Considering two different passwords, a \textit{start} password and a \textit{target} password, we show that we can use PassFlow to interpolate in the latent space from \textit{start} password to the \textit{target} password.
Algorithm~\ref{alg:interpolation} shows the pseudocode of our implementation of interpolation. We first preprocess the \textit{start} and \textit{target} passwords (see Section~\ref{sec:dataset}) and obtain their latent representation $z_1$ and $z_2$ respectively. Given a predefined number of interpolation \textit{steps} to move from \textit{start} to \textit{target} password, we compute $\delta$ (the step size in the latent space) by taking the difference between $z_2$ and $z_1$ and dividing that value by the desired number of \textit{steps}.
Iterating over the number of \textit{steps}, we generate intermediate points $i$ by adding to $z_1$ the step size $\delta$ times the current step number.

\begin{algorithm}
\DontPrintSemicolon
  
  \KwInput{Str: $start$, Str: $target$, Int: $steps$}
  \KwOutput{Set: $interpolations$}
  \KwData{Set: $interpolations$}

    $x_1 \leftarrow preprocess(start)$ \;
    $x_2 \leftarrow preprocess(target)$ \;
    
    $z_1 \leftarrow f_\theta(x_1)$ \;
    $z_2 \leftarrow f_\theta(x_2)$ \;
    
    $\delta \leftarrow (z_2 - z_1) / steps$ \;
    \For{$j \gets 0$  \KwTo $ steps$}{
         $i \leftarrow  f_\theta^{-1}(z_1 + \delta * j)$ \;
   		 $interpolations \leftarrow interpolations \cup i$ \;
    }
\caption{Interpolation}
\label{alg:interpolation}
\end{algorithm}

\subsection{Dataset and Model Parameters}\label{sec:dataset}

The RockYou dataset~\cite{rockyou_dataset} contains $\sim32,5M$ passwords. To directly compare our flow-based model with previous work, we use the same setting of Hitaj et al.~\cite{passGAN} and Pasquini et al.~\cite{pasquini_representation_learning} by training PassFlow on passwords of length 10 or less. This results in a total of roughly 29,5 millions passwords from the whole RockYou dataset. This reduced dataset is split in a 80/20 ratio as done also by Hitaj et al.~\cite{passGAN}, where 80\% is for training and the remaining 20\% for testing. 
Out of the 80\% selected for the training ($\sim$23.5 million passwords) we randomly sample 300,000 instances that we use to train our model.
The remaining 20\% that we use as the test set, accounts for approximately 6 million passwords of length 10  or less. This set is cleaned the same way as~\cite{passGAN} by removing duplicates and intersection with the training set, resulting in approximately 1.94 million unique instances which we use to evaluate and compare the password generation performance of PassFlow.

The PassFlow architecture uses 18 coupling layers and all the binary masks used have m=1 (see Section~\ref{sec:coupling_layer}). Each coupling layer's $s$ and $t$ functions are based on ResNet~\cite{resnet} architecture  containing 2 residual blocks with a hidden size of 256 units.
PassFlow was trained over 400 epochs using Adam~\cite{kingma2014method} optimizer with a batch size of 512 and learning rate equal to 0,001. We pick the best performing epoch for our password generation task. Before feeding the data for training we convert the passwords in feature vectors that contain their numerical representation and then we normalize by the size of the alphabet. 
\section{Evaluation}\label{sec:evaluation}

\begin{table*}[t]
\centering
\caption{The \% of matched passwords by GANs from Hitaj et al.~\cite{passGAN} and Pasquini et al., CWAE from Pasquini et al.~\cite{pasquini_representation_learning} and \textit{PassFlow}, over the RockYou~\cite{rockyou_dataset} test set ($\sim$1.94 million).}
\label{table:results_comparisons}

\begin{tabular}{l ccccc}
\toprule
 & \multicolumn{5}{c}{Number of Guesses}\\
{Method} & $10^4$ & $10^5$ &$10^6$ & $10^7$ & $10^8$ \\ 
\cmidrule(r){1-1} \cmidrule(l){2-6}
PassGAN, Hitaj \emph{et al.}~\cite{passGAN} & 0.01 & 0.05 & 0.38	& 2.04	& 6.63 \\ 
GAN, Pasquini \emph{et al.} ~\cite{pasquini_representation_learning} & - & - & -	& -	& 9.51 \\ 
CWAE, Pasquini \emph{et al.}~\cite{pasquini_representation_learning}  & 0.00 & 0.00 &  0.05	& 0.42	& 3.06 \\ 
PassFlow-Static  & 0.00 & 0.01 &  0.10	& 0.82	& 3.95 \\ 
PassFlow-Dynamic  & 0.01 & 0.12 &  0.59 & 2.60 & 8.08 \\ 
PassFlow-Dynamic+GS  & 0.01 & 0.13 &  0.78 & 3.37 & 9.92 \\ 
\bottomrule
\end{tabular}
\end{table*}

\subsection{Password Guessing Performance}

Table~\ref{table:results_comparisons} presents a comparison among the password guessing performance of our flow-based generative model, with and without dynamic sampling and dynamic sampling with Gaussian Smoothing (PassFlow-Static, PassFlow-Dynamic, PassFlow-Dynamic+GS respectively) and that of prior art. 
In particular, we compare against the GAN based approaches of Pasquini et al.~\cite{pasquini_representation_learning} and of Hitaj et al.~\cite{passGAN} (PassGAN), and against the Context Wasserstein Autoencoder (CWAE) deep latent variable model of Pasquini et al.~\cite{pasquini_representation_learning}. For each approach, the table shows the percentage of samples generated that match a password present in the RockYou test set (see Section~\ref{sec:dataset}). Care was taken to appropriately clean the test set by removing any intersection between the training and test set, as well as removing any duplicate passwords in the test set. This is done to provide a precise evaluation of the generalization performance of the models, excluding potential overfitting artifacts. 

As Table~\ref{table:results_comparisons} shows, the performance of PassFlow-Static is superior to that of comparable deep latent variable models such as CWAE for all sample sizes. Nevertheless, the performance of PassFlow-Static cannot match that of GANs, trailing both Hitaj et al. and Pasquini et al.'s approaches on all sample sizes. When we consider the performance of PassFlow-Dynamic, we can see that our approach outperforms both CWAE and PassGAN on all sample sizes, often by a considerable margin, while remaining competitive with the GAN of Pasquini et al. on most sample sizes. However, when we consider the performance of PassFlow-Dynamic+GS, we can see that our approach outperforms all previous password guessing approaches on all sample sizes.

We believe that these results are remarkable for a number of reasons: (1) while GAN architectures have seen extensive research and fine-tuning, flow-based models are still in their infancy and are only now beginning to attract more attention from the research community~\cite{realNVP,glowINN,MaCow,NEURIPS2019_5d0d5594, pmlr-v97-ho19a, pmlr-v119-ping20a, 9054484}. It seems reasonable that, in the near future, flow-based architectures will be improved and fine-tuned similarly to what happened with GANs. We would then expect to see similar improvements for flow-based architectures as we see between PassGAN and the GAN of Pasquini et al.; (2) All PassFlow results in Table~\ref{table:results_comparisons} are obtained with a training set comprised of only $300K$ password samples, which is \emph{2 orders of magnitude smaller} than the training set used by the other approaches ($\sim23.5$ million samples). This proves the ability of flow-based architectures to correctly model the dataset distribution and generalize well with much less information than other types of architectures.

Table~\ref{table:results_comparisons_latent} provides a detailed comparison between the four latent space models CWAE, PassFlow-Static, PassFlow-Dynamic and PassFlow-Dynamic+GS, with a particular focus on the number of unique passwords generated. As we can notice, the margin by which PassFlow-Static outperforms CWAE decreases proportionally to the increase in the difference of unique passwords generated. 
Considering $10^8$ generated passwords, CWAE samples $\sim10\%$ more unique passwords than PassFlow-Static. However, PassFlow-Static is able to outperform CWAE on this sample size too. This increasing difference in the number of unique passwords generated is related to the dimensionality of the latent space of the two models. Since CWAE is based on an AutoEncoder architecture, the dimensionality of the latent space can be varied arbitrarily and is not bound by the dimensionality of the data space. In~\cite{pasquini_representation_learning}, the authors chose 128 dimensions to represent data in the latent space. However, with flow architectures we cannot choose an arbitrary dimensionality for the latent space, as it is tied to the dimensionality of the data space. In our evaluation, similar to prior work on the password guessing domain~\cite{passGAN, pasquini_representation_learning}, we use passwords with max length of $10$ characters to train and test our models, and therefore the latent space is also bound to $10$ dimensions. Consequently, we have a much higher probability of generating duplicate data points from our 10-dimensional Gaussian compared to the 128-dimensional Gaussian in~\cite{pasquini_representation_learning}. This issue is exacerbated by the transition from a continuous latent space to a descrete data space, as previously discussed in Section~\ref{sec:gs}.

On the other hand, the number of unique samples generated by our dynamic sampling approach are even lower than that of PassFlow-Static. This is expected, as with dynamic sampling we are altering the prior distribution of the latent space by conditioning on the set of successfully matched passwords. Effectively, we are altering the mean and variance of the D-dimensional Gaussian used for latent space sampling, restricting the regions of the manifold that we explore to those close to already matched passwords. Nonetheless, since we are exploring regions of the latent space with high probability density (i.e., likely real passwords), we are able to match a much larger number of passwords in the test set and heavily outperform CWAE. 
Finally, PassFlow-Dynamic+GS, which adds data-space Gaussian Smoothing to the generation process, greatly improves both the amount of unique samples generated and the number of matched passwords for all sample sizes, outperforming both CWAEs and GANs.

\vspace{.8em}
\par\noindent\textbf{{Non-matched samples}} \\
Beside the number of matched passwords generated by the model, an important metric to consider is how well non-matched passwords follow the structure and patterns of human-generated passwords. Indeed, even if non-matched passwords were not part of the specific test set used in our experiments, if they follow a human-generated distribution they could be a match in other test sets. On the other hand, if non-matched passwords generated by PassFlow show very little to no structure, then this would cast a doubt on the capability of the model to generalize well and would highlight limitations in its ability to consistently generate high-quality guesses.

In Table~\ref{tab:non_matched_passwords} we report a few password samples generated using PassFlow that were not matched on the RockYou test set. We notice that the samples generated resemble human-like passwords, despite not matching any password in the test set. This provides further evidence that that PassFlow is able to precisely model the space of human generated passwords.

\begin{table*}[t]
\centering
\caption{The number of unique and matched passwords by CWAE from Pasquini \emph{et al.} ~\cite{pasquini_representation_learning}, \textit{PassFlow} with and without dynamic sampling and PassFlow with dynamic sampling with Gaussian smoothing over the RockYou~\cite{rockyou_dataset} test set.}

\label{table:results_comparisons_latent}
\begin{tabular}{lrrrrrrrr}
\toprule
& \multicolumn{2}{r}{\bfseries CWAE~\cite{pasquini_representation_learning}} &
\multicolumn{2}{r}{\bfseries PassFlow-Static} &
\multicolumn{2}{r}{\bfseries PassFlow-Dynamic} &
\multicolumn{2}{r}{\bfseries PassFlow-Dynamic+GS} \\
\cmidrule(l){2-3} \cmidrule(l){4-5}  \cmidrule(l){6-7} \cmidrule(l){8-9}
Guesses &Unique &Matched &Unique  &Matched &Unique &Matched &Unique &Matched  \\
\midrule
$10^4$ & 9,942 & 9 & 9,986 & 31 & 8,753 & 252 & 9,053 & 285 \\
$10^5$ & 99,055 & 98 & 99,506 & 212 & 77,171 & 2,221 & 93,605 & 2635  \\
$10^6$ & 987,845 & 960 & 989,707 & 2,056 & 717,787 & 11,528 & 973,248 & 15,037  \\
$10^7$ & 9,832,738 & 8,411 & 9,662,309 & 16,026 & 5,520,518 & 50,510 & 9,748,457 & 65,067 \\
$10^8$ & 96,979,169 & 60,630 & 87,902,573 & 76,787 & 58,997,803 & 157,023 & 89,214, 314 & 191,114 \\
\bottomrule
\end{tabular}
\end{table*}

\begin{table}[t]
\centering
\caption{Samples generated by PassFlow that did not match any password in the testing set.}
\label{tab:non_matched_passwords}
\begin{tabular}{l l l l}
\hline
{\tt gaslin} & {\tt gala8a} & {\tt 7atam} & {\tt jezes1} \\
{\tt kaoni1} & {\tt  pa5ase } & {\tt  ra8ona } & {\tt  9alalm } \\
{\tt s1cila } & {\tt  deyele } & {\tt  ragena } & {\tt seytcy } \\
{\tt 4an1sa } & {\tt  lema0 } & {\tt  9anl2a } & {\tt  lasata } \\
{\tt manona } & {\tt  vanroe } & {\tt  leA191 } & {\tt  lamaon } \\
{\tt raj19 } & {\tt  caoao } & {\tt  sanan } & {\tt  mapali } \\
{\tt sav1ii} & {\tt  hetara } & {\tt  2ebeta } & {\tt  janeo} \\
{\tt  gagano } & {\tt  9aneta }& {\tt  sakni } & {\tt  ve2o0a } \\
{\tt  kasa9 } & {\tt  magata } & {\tt  dalria } & {\tt  sara4}\\\hline
\end{tabular}
\end{table}


\subsection{Smoothness of the Latent Space}
\label{sec:eval_smoothness}

\begin{table}[t]
\centering
\caption{The first 10 unique passwords obtained with different values of $\sigma$ starting from the pivot string ''\textit{jimmy91}"}
\label{table:smoothness_latent_space}

\begin{tabular}{cccc}
\cmidrule(r){1-4}
$\sigma = 0.05$ & $\sigma = 0.08$ & $\sigma = 0.10$ & $\sigma = 0.15$ \\
\cmidrule(r){1-4}
ji3myte	& jimtym1	& jim3h3i	& p10td3i  \\
jimmd3i	& jimtdoe	& jimmdsl  & ji3mym1 \\
jimmy31	& ji334te	& vi39dno & pi33yme \\
jimmy3i & vimtyc1	& pim3dte & jimm4c1 \\
vimmy91	& jim0yte	& jimmy81 & jimt4se \\
jimmyc1 & vi33y9i   & vimtd0o & gi9349i \\
jimmyte & jimtdte   & pimmyme & vimmdso \\ 
jimmyci & jim3431   & j10ty9i & jimtd0o \\ 
jim3dmi & jimt4me   & vimmhrn & ji33d00 \\ 
jimtyme & vimmd91   & jimmd3i & jim3yno \\ 

\cmidrule(r){1-4}
\end{tabular}
\end{table}

The mapping function $f_\theta$ learned by our model and the factorized structure of the prior $p_z$ enforce a smooth distribution in the latent space, as well as imposing locality constraints on latent points that share semantic relations in the data space. These attributes of the latent space result in a number of interesting properties and enable meaningful manipulation of latent space representations of datapoints (i.e. passwords). In this section, we analyze the geometry of the latent space learned by our model and prove the above-mentioned smoothness and locality properties.

The locality property of the latent space implies that similar classes of passwords (i.e passwords with similar structures and patterns) are organized in related areas of the latent space. This property can be used to introduce biases in the password generation process, forcing the model to explore regions of interest in the space. We can generate instances of passwords belonging to a specific class by bounding the sampling to specific subspaces of the latent space. Table~\ref{table:smoothness_latent_space} provides an example of such bounded sampling. In this example, we sample a set of latent points in the neighborhood of the latent representation of the string ``jimmy91'', which is used as a pivot to guide the sampling process. We parametrize the sampling based on the standard deviation of the Gaussian used for the prior $p_z$, starting with $\sigma = 0.05$. As expected, datapoints that reside closer to the pivot in the latent space correspond to passwords that share a strong semantic relation with it. This can be particularly observed in column 1, where most of the sampled passwords retain a structure that is almost identical to that of the pivot, while producing meaningful variations that are coherent with the underlying distribution (e.g., ``jimmy3i'', ``vimmy91''). Increasing $\sigma$ alters the structure of the generated samples compared to the pivot, but still maintains strong similarities with it (i.e similar mixes of letters and numbers; letters with similar sounds that people use interchangeably in passwords), highlighting once again the strong semantic organization of the latent space.

\begin{figure*}[tb]
	    \centering
	    \begin{subfigure}{.47\textwidth}
	        \centering
                \includegraphics[width=.9\columnwidth]{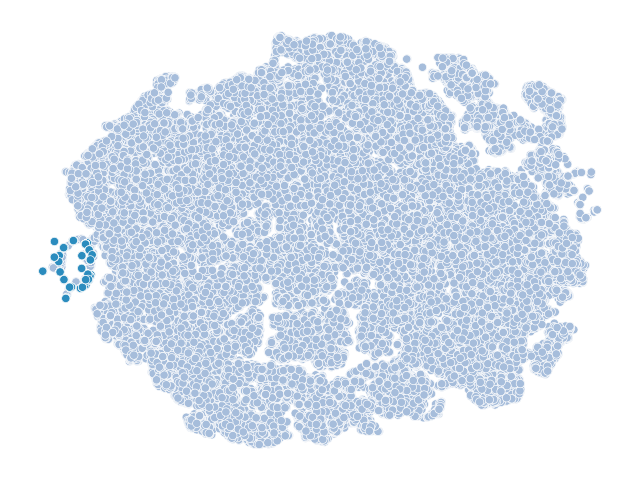}
                
            \caption{Latent space around \textit{"jaram"}. Some of the samples include: jara0, jara3, jala9, jaoat, jaoa0.}
            \label{fig:latent_jaram}
        \end{subfigure}
        \hfill
        \begin{subfigure}{.47\textwidth}
            \centering
            \includegraphics[width=.9\columnwidth]{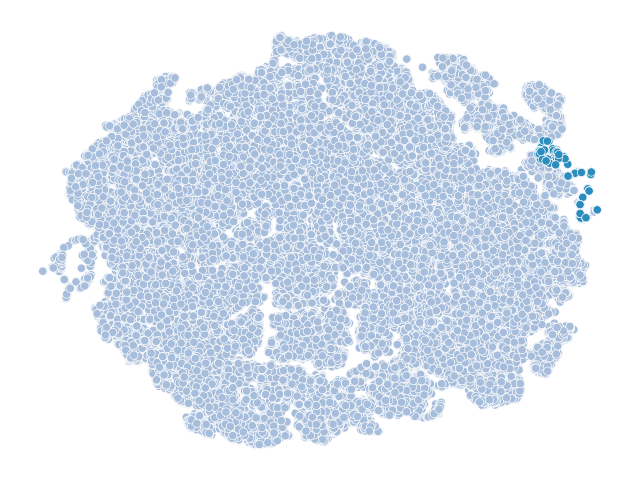}
            
            \caption{Latent space around \textit{"royal"}. Some of the samples include: ro4al, ro5al, lohal, rocal, ooyal.}
            \label{fig:latent_royal}
        \end{subfigure}
        \caption{Projection of a sample of latent space points in the neighborhood of the passwords ``jaram'' (to the left) and ``royal'' (to the right) over the latent space learned by our model. Image generated using TSNE~\cite{van2008visualizing}.}%
        \label{fig:latent_space}%
        
	\end{figure*}

Figure~\ref{fig:latent_space} provides a visual representation of the semantic structure of the latent space of PassFlow. This figure shows a two-dimensional projection of the latent space learned by our model (light blue background) using the TSNE tool~\cite{van2008visualizing}.
We generate the latent representation of the two passwords ``jaram'' and ``royal'', sample some latent points in their vicinity, and cast these obtained latent points on top of projected latent space (dark blue points; Figure~\ref{fig:latent_jaram} for ``jaram'' vicinity, Figure~\ref{fig:latent_royal} for ``royal'' vicinity). Some of the latent points sampled in the vicinity of ``jaram'' include jaoa0, jana0, jalat; some of the latent points in the vicinity of ``royal'' include ro4al, lohal, ro8al. As we can see in the figure, all these syntactically similar passwords are mapped to regions of the latent space that share a strong spatial correlation.

\begin{figure*}[tb]
    \centering
    \includegraphics[width=.9\textwidth]{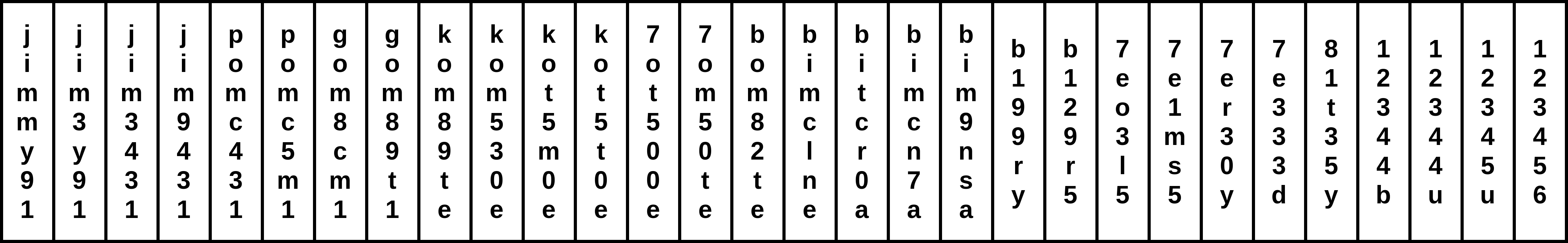}
    \caption{Interpolation in the latent space between the latent representation of ``jimmy91'' and ``123456'', mapped back to the password space. Left-to-right.}
    \label{fig:interpolation}
\end{figure*}

In Figure~\ref{fig:interpolation} we provide evidence for the smoothness of the latent space learned by PassFlow. If the latent space learned by our model is indeed smooth, moving in a bounded region around a latent point with high probability density, should yield other latent points with similarly high probability density. Effectively, this means that moving in the vicinity of the latent representation of a real password, should yield latent points corresponding to similarly realistic passwords. In particular, if the above statement is true, then recursively sampling in the vicinity of a latent point starting from a region with high probability density should yield points that all share similarly high probability density. We can apply this concept to perform \emph{interpolation} between passwords in the latent space. In Figure~\ref{fig:interpolation}, we interpolate between the passwords (1) ``jimmy91'' and (2) ``123456'' by moving in the latent space from the latent point corresponding to (1), to the latent point corresponding to (2). While interpolating in the latent space, we sample the intermediate points along the path and map them back to the data space of passwords. As we can see, most of the samples generated through interpolation present structure and patterns typical of human-generated passwords, i.e. they are all \emph{high density points} in the latent space, which demonstrate its smoothness. Moreover, consecutive samples share similar characteristics, as expected. While some samples are not obviously human-like (i.e., they don't appear to be alterations of known words/names), it's important to remember that by interpolating along a line from source to destination password we're essentially forcing the model to generate guesses from a set of fixed points in the latent space, without considering their probability density. While sampling in a bounded region around an initial high-density point in a smooth latent space yields similarly high-density points, the further we move from the origin the more we increase the chance of sampling lower density points. Therefore, it is expected that some of the interpolated samples are not immediately recognizable as altered words or names. However, they still retain similar structure and patterns to those of other, human-like passwords.
Interpolating between passwords can be useful in case we have some previous knowledge of the password we are trying to guess (e.g., the password contains the name of a person, plus another specific characteristic, such as it contains numbers). Using interpolation in the latent space, and using the previous knowledge, we can then generate high probability guesses on the potential passwords. 

\begin{figure}
    \centering
    \includegraphics[width=.85\columnwidth]{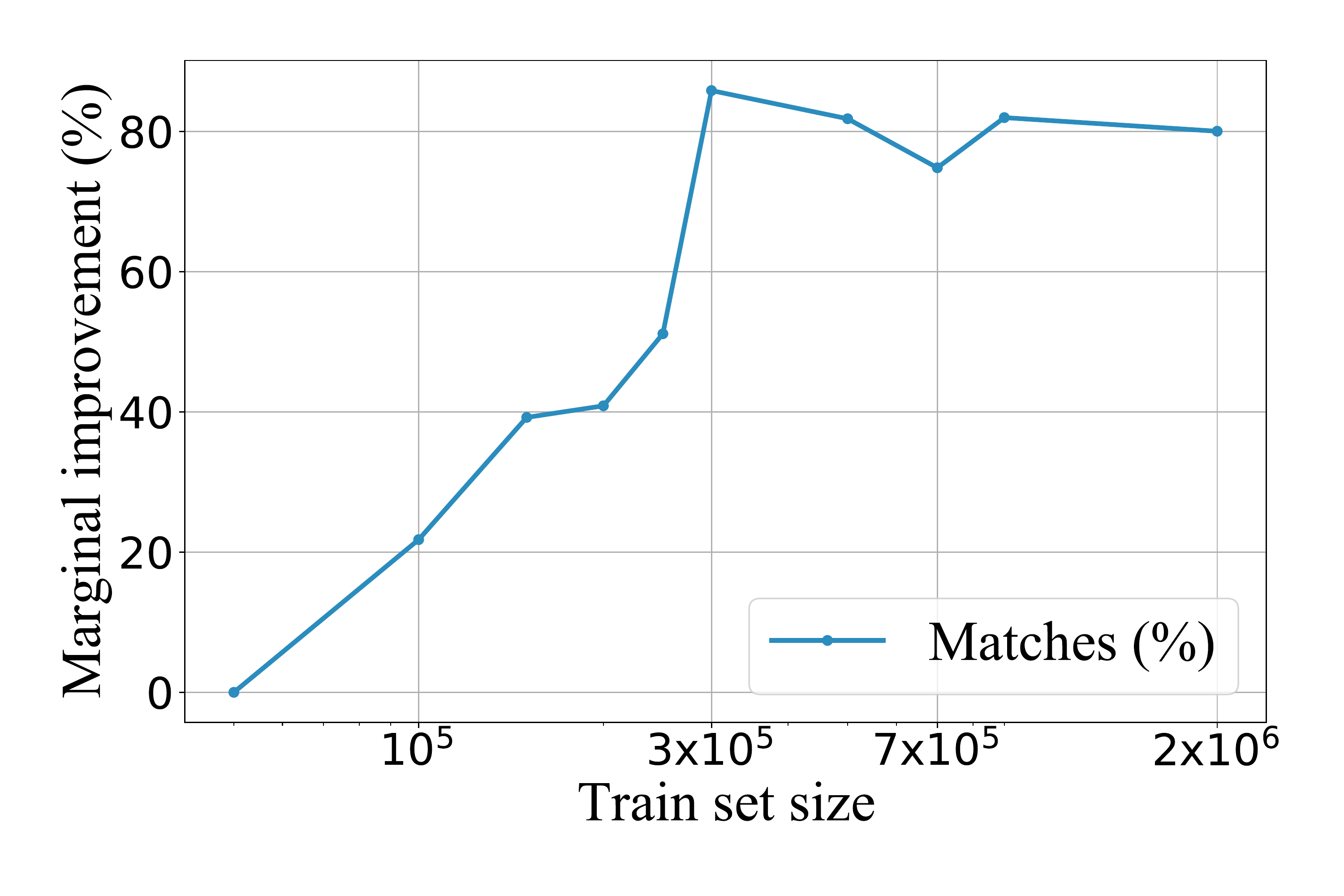} 
    \caption{The marginal performance improvement, with different models trained on varying dataset sizes. The improvement is calculated against the performance of a baseline PassFlow model trained with $50K$ samples.}
    \label{fig:dataset_marginal}
\end{figure}

\subsection{Effect of different Masking patterns}
\label{sec:masking_patterns_evaluation}
The ability of the model to learn local correlations between the characters of a password hinges on the masking strategy used in the coupling layers of the network. Consequently, the choice of masking strategy has a considerable impact on the overall performance of the architecture. In our evaluation, we implemented and tested three different types of masking patterns:  \emph{horizontal} masking, \emph{char-run 1} masking and \emph{char-run 2} masking. Specifically, we define horizontal masking as a \emph{binary mask} $b \in \{0,1\}^D$, comprised of $D / 2$ consecutive zeroes followed by $D/2$ consecutive ones, where $D$ is the password length. Effectively, this masking strategy splits passwords in half, conditioning the first half of the password on the second half (and vice-versa for the layers with the inverted mask, see Section~\ref{sec:coupling_layer}). We define char-run $m$ masking as a binary mask $b \in \{0,1\}^D$ with runs of \emph{m} alternating consecutive zeroes and ones. For instance, \emph{char-run 1} masking is a binary mask of length $D$ of alternating zero and ones, while \emph{char-run 2} is a binary mask of alternating pairs of zeroes and pairs of ones (e.g., $00110011$).

We trained PassFlow with horizontal masking, char-run masking with $m = 1$ and char-run with $m = 2$, and evaluated its sampling performance. As shown in Table~\ref{table:results_comparisons_maskings}, best results were obtained when using the \emph{char-run} masking pattern with $m = 1$. All the results presented in these sections are obtained using this masking pattern.

\begin{table}[t]
\centering
\caption{The number of matched passwords obtained by PassFlow trained with three different Masking strategies}

\label{table:results_comparisons_maskings}
\begin{tabular}{lrrr}
\toprule
& \multicolumn{1}{r}{\bfseries Horizontal} &
\multicolumn{1}{r}{\bfseries Char-run 2} &
\multicolumn{1}{r}{\bfseries Char-run 1} \\
\cmidrule(l){2-4}
Guesses &Matched  &Matched &Matched   \\
\midrule
$10^4$ & 10 & 15 & 31 \\
$10^5$ & 120 & 148 & 212  \\
$10^6$ & 1,091 & 1,449 & 2,056  \\
$10^7$ & 9,832 & 11,833 & 16,026 \\
$10^8$ & 60,248 & 59,184 & 76,787  \\
\bottomrule
\end{tabular}
\end{table}

\subsection{Effects of Training Set Size and $\phi$ on Performance}
\label{sec:phi_train_size_performance}
As discussed previously, PassFlow obtains competitive performance with prior art while using a training set that is comparatively orders of magnitude smaller. Figure \ref{fig:dataset_marginal} plots the marginal performance improvement in matches found over the test set, when training different PassFlow models with varying dataset sizes.
Regardless of the specific training set size used, all models are evaluated on $20\%$ of the complete RockYou dataset. We calculate the marginal performance improvement compared to the performance of PassFlow trained with $50K$ samples, which we use as baseline. 
 
We can see that there is an initial sharp increase in performance, $\sim40$ percentage points, when going from $50K$ training samples to $150K$ training samples, and then again a $\sim45$ percentage points increase from $200K$ to $300K$ samples, where performance peaks. 
The marginal improvement then reaches a plateau where the performance of the model does not increase with increasing size of the training set. These results indicate that flow-based models are able to generalize exceptionally well and precisely approximate the probability density of the password domain. However, they also point to a limited ability of our model to disentangle features in the latent space. We hypothesize that this behavior is attributable to the limited dimensionality of the latent spacce and of the intermediate representations that follow each coupling layer, which in flow models is always equal to the dimensionality of the data space. Some novel works on flow networks have analyzed this limitation of these architectures and have proposed potential solutions~\cite{chen2020vflow}. We reserve a deeper analysis of the advantages and disadvantages of greater latent dimensionality for our future work.

\begin{figure}
    \centering
    \includegraphics[width=.85\columnwidth]{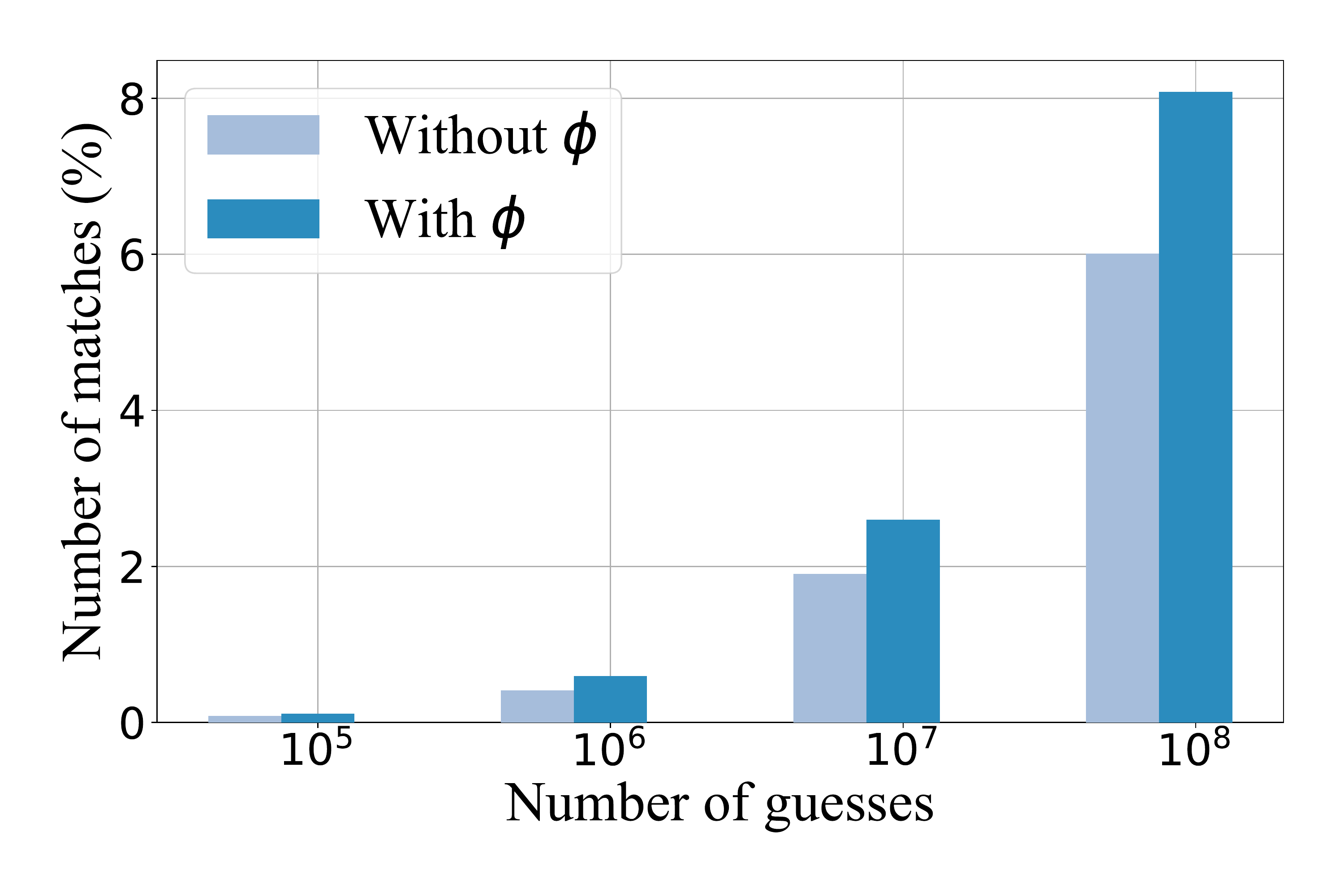}
    \caption{The performance improvement with and without $\phi$ for varying sample sizes.}
    \label{fig:phi}
\end{figure}

Figure~\ref{fig:phi} shows a bar plot illustrating the difference in performance in passwords matched by PassFlow-Dynamic with and without the penalization function $\phi$. The results labeled ``without $\phi$'' are obtained by setting $\phi=1$ in Equation~\ref{eq:ds} (i.e., uniform weighting for all matched samples, regardless of history). 
As we can see, the penalization factor provides a substantial performance improvement across all sample sizes, becoming more significant for larger sample sizes. For instance, when generating $10^7$ guesses with PassFlow-Dynamic, $\phi$ increases the number of matched passwords by 1.78 percentage points ($0.82\%$ without, $2.60\%$ with). 
When generating $10^8$ samples, the increase in performance when using $\phi$ grows to 4.13 percentage points ($3.95\%$ without, $8.08\%$ with).
These results highlight the importance of the penalization function in the definition of the posterior probability in Equation~\ref{eq:ds}. Effectively, the penalization function alters the posterior probability to discourage stagnation in regions of the latent space that have already been sufficiently explored, promoting search in other areas of the latent space with high probability density. In Section~\ref{app:phi}, we discuss the implementation of $\phi$ used in our experiments, as well as other potentially suitable functions.

\section{Related Work}\label{sec:related_work}

In the past decade, research in machine learning and deep neural networks has seen astonishing progress, with significant impact on various domains of computer science such as image recognition~\cite{Simonyan14verydeep, He_2016_CVPR, Chollet2017XceptionDL}, image generation~\cite{invertible_residual_networks, irevnet}, speech recognition~\cite{speech2, speech3}, natural language processing~\cite{nlp1, nlp2} and many more. These improvements in ML techniques have also led to novel applications in the field of cybersecurity, where ML has been used to improve numerous aspects in this domain, such as intrusion detection~\cite{intrusion_detection, malphase}, ransomware detection~\cite{continella_shieldfs, encod_ndss, mehnaz_rwguard}, advanced persistent threats detection~\cite{unicorn}, but also introduced novel threats and vulnerabilities to cybersecurity systems~\cite{naked_sun, evasion_watermarking, passGAN, pasquini_representation_learning}.

Closely related to the scope of this paper, i.e the domain of password guessing, alongside classic rule-based tools such as HashCat~\cite{hashcat} or JTR~\cite{jtr}, Weir~\textit{et al.}~\cite{probabilistic_grammars}  proposed a technique to automatically create a context-free grammar that allows to model the distribution of human-chosen passwords, based on previously disclosed password datasets. This context-free grammar allows the generation of word-mangling rules, and from them, high-probability password guesses that can be used in password cracking. Noting that artificial neural networks are very good distribution approximators, Melicher~\textit{et al.}~\cite{modelling_password} exploit these architectures to model the resistance of text-based passwords to guessing attacks, and explore how different architectures and training methods impact the effectiveness of the network. Moreover, Melicher~\textit{et al.}~\cite{modelling_password} demonstrated that neural networks outperform context-free grammars and Markov models (such as~\cite{jtr_markov}) in password guessing.

With the introduction of generative models, various approaches achieving state-of-the-art performance on standard datasets have been proposed. In the following sections, we discuss in detail the main related works on password guessing based on these architectures.

\subsection{PassGAN~\cite{passGAN}}
Generative adversarial networks (GANs)~\cite{goodfellow2014generative} are a type of adversarial learning scheme to train a deep generative model that is able to generate samples from the same distribution as the one of the training set $I=\{x_{1},x_{2},\dots, x_{n},\}$. The learning scheme typically consists of 2 components trained simultaneously: a \textit{generator} $G$ and a \textit{discriminator} $D$. The generator $G$ aims at generating samples from the desired distribution by relying on the discriminative ability of $D$, which is trained to distinguish between samples coming from the distribution of training set $I$ and the samples generated by $G$.
Formally, considering an input noise distribution (e.g, normal distribution), this learning scheme is represented as a \textit{minimax} optimization problem as shown in Equation~\ref{eq:gans}:

\begin{equation}\label{eq:gans}
    \min_{\theta_{G}} \max_{\theta_{D}} \sum_{i=1}^{n} \log f(x_{i}; \theta_{D}) + \sum_{j=1}^{n} \log(1-f(g(z_{j};\theta_{G}); \theta_{D}))
\end{equation}
where $f(x_{i}; \theta_{D})$ represents the discriminator $D$
and $(g(z_{j};\theta_{G})$ represents the generator $G$. The learning is considered complete when the discriminator $D$ is not able to distinguish anymore between synthetic samples generated from $G$ and the real samples belonging to $I$.

Hitaj~\textit{et al.}~\cite{passGAN} introduce PassGAN, that makes use of a variation of the initial learning scheme by Goodfellow~\textit{et al.}~\cite{goodfellow2014generative}. Specifically, they use Wasserstein-GAN with gradient penalty~\cite{Gulrajani2017ImprovedTO} where the generator $G$ is based on residual neural network architecture~\cite{He_2016_CVPR}. This generator is used to estimate the password distribution over the RockYou dataset~\cite{rockyou_dataset}, considering passwords with 10 or fewer characters. The authors train PassGAN on 23 million passwords, $\sim$10 million of which are unique. 
The performance of PassGAN is generally competitive with that of rule-based approaches such as HashCat~\cite{hashcat} or JTR~\cite{jtr}. However, being the first application of an unsupervised learning scheme in the domain of password guessing, PassGan oftentimes requires the generation of larger number of passwords to obtain the same number of matches as the above-mentioned tools on the RockYou test set.
The limitations of PassGAN come due to the fact that GANs do not allow for explicit estimation of the probability density of the training data, but instead they approximate the stochastic procedure that generates the data~\cite{mohamed2017learning}. This phenomena combined with the challenging optimization problem used to train GANs, which in itself introduces multiple issues (e.g, mode collapse, posterior collapse, vanishing gradients, training instability), hinder the ability of PassGAN to efficiently generate a large number of diverse passwords (thus needing a larger number of generated samples to match the performance of HashCat~\cite{hashcat} or JTR~\cite{jtr}).

\subsection{GAN by Pasquini \textit{et al.}~\cite{pasquini_representation_learning}}\label{sec:gan_pasquini}

Pasquini \textit{et al.}~\cite{pasquini_representation_learning} tackle PassGANs issues by introducing a form of stochastic smoothing over the representation of the strings in the train set. Effectively, the authors alter the one-hot encoded password representation, which is also used in PassGAN, with a small additive noise sampled from a uniform distribution. This additive noise allows the GAN model to perform up to 30 times more training iterations than PassGAN, without suffering from training instability, keeping the general GAN framework similar. In their GAN framework, Pasquini \textit{et al.} use a deeper model architecture for both the generator and the discriminator, by substituting the residual blocks used in PassGAN with deeper residual bottleneck blocks~\cite{He_2016_CVPR}. Moreover they introduce batch-normalization in the generator, which coupled with skip connections of residual blocks allow for even deeper architectures. With this improvement, they are able to outperform PassGAN by generating more matches in the RockYou test set.

\subsection{Deep Latent Variable Models for Password Guessing}\label{sec:ae_pasquini}
Another generative approach proposed by Pasquini~\textit{et al.}~\cite{pasquini_representation_learning} is the use of autoencoders, more specifically Wasserstein autoencoder (WAE)~\cite{WAE} with moment matching regularization. Unlike GANs, autoencoders allow for sampling from a latent space, since they provide an explicit latent space organized according to a given a prior distribution.
The autoencoder architecture is composed of 2 components; an \textit{Encoder} that learns to transform input data into a structured latent representation, and a \textit{Decoder}, which takes as input this latent representation and reconstructs the initial input. 

In~\cite{pasquini_representation_learning}, the authors train the autoencoder as a context one (CWAE), by following~\cite{pathakCVPR16context}, in order to improve regularization. 
The training is performed as follows: Given a password $x_{i}$, the encoder is fed a noisy version $\Tilde{x_{i}}$. The noisy version is obtained by removing certain characters from $x_{i}$ with a probability $p = \frac{\epsilon}{|x_{i}|}$, where $|x_{i}|$ is the length of the password and $\epsilon$ is a fixed hyperparameter.
The decoder is trained to reconstruct the original password $x_{i}$, meaning that the model is trained to estimate the missing characters from the context given by the encoded version of $\Tilde{x_{i}}$.
The autoencoders are optimized using a function that provides a lower bound on the expectation of a sample, thus being able to learn an approximate latent variable inference. On the other hand, flow architectures like \textit{PassFlow} are optimized using exact log-likelihood, thus being able to provide exact latent variable inference.
This inherent property of our flow-based generative approach allows us to heavily outperform deep latent variable models in the password guessing task as shown on Table~\ref{table:results_comparisons_latent}.

\section{Limitations and Future Work}

While our flow-based approach shows promising results, there are a few limitations that require further research and need to be addressed. 
Using the current PassFlow architecture we are unable to perform conditional guessing operations. Conditional guessing refers to the process of guessing the most likely password given a partial password. For instance, given the password ``jimmy**'', guess the complete high probability password ``jimmy91''. This limitation is directly inherited from the traditional generative flow architectures which are unable to perform this type of operation. 

As future work, we intend to analyzing the applicability of \emph{conditional} normalizing flows~\cite{winkler2019learning} to address the limitations of our architecture with respect to conditional guessing. Conditional flows are an extension of normalizing flows where the mapping between probability density and output space is conditioned on an additional input y. Conditional architectures enable the direct modeling of conditional probability densities, which will allow us to perform conditional password guessing. Finally, we plan to study the effects of different penalization functions in order to improve the performance of dynamic sampling.

\section{Conclusions}\label{sec:conclusions}

This work presented PassFlow, a flow-based generative approach to the password guessing task. PassFlow pushes further the state-of-the-art in the password guessing task by outperforming prior work based on deep latent variable models and GANs, all while using a training set that is orders of magnitude smaller than prior art. 
We performed an extensive qualitative evaluation of the passwords generated with PassFlow, as well as a thorough analysis of the structure of the latent space learned by our model. We showed that flow-based architectures are able to accurately model the original password distribution and learn a smooth, well-structured latent space. These properties enable PassFlow to perform complex operations such as interpolation between passwords and precise exploration of dense regions of the latent space. We experimentally proved these properties and proposed dynamic sampling with penalization function, an advanced sampling technique that exploits the structure of our latent space.
Furthermore, non matched samples generated by PassFlow closely resemble human-like passwords, providing further evidence of the robust generalization performance of the model.


\balance
\bibliographystyle{IEEEtranS}
\bibliography{main}

\end{document}